\documentclass[superscriptaddress,pra,aps,twocolumn,10pt,longbibliography]{revtex4-1}

\usepackage{amsfonts}
\usepackage{amssymb}
\usepackage{amsmath}
\usepackage{graphicx}
\usepackage{verbatim}
\usepackage{hyperref}
\usepackage{braket}
\usepackage{float}
\usepackage{grffile}
\usepackage[english]{babel}
\usepackage{natbib}
\usepackage[pdftex]{color}
\usepackage[dvipsnames]{xcolor}
\usepackage{soul}

\hypersetup{
    colorlinks=true,       
    linkcolor=cyan,          
    citecolor=magenta,        
    filecolor=magenta,      
    urlcolor=cyan,           
    runcolor=cyan
}

\begin{document}
\title{Master Equation Emulation and Coherence Preservation with Classical Control of a Superconducting Qubit}

\author{Evangelos Vlachos}
\affiliation{Department of Physics \& Astronomy, Dornsife College of Letters, Arts, \& Sciences, University of Southern California, Los Angeles, CA 90089, USA}
\affiliation{Center for Quantum Information Science \& Technology, University of
Southern California, Los Angeles, CA 90089, USA}
\author{Haimeng Zhang}
\affiliation{Department of Electrical Engineering, Viterbi School of Engineering, University of Southern California, Los Angeles, CA 90089, USA}
\affiliation{Center for Quantum Information Science \& Technology, University of
Southern California, Los Angeles, CA 90089, USA}
\author{Vivek Maurya}
\affiliation{Department of Physics \& Astronomy, Dornsife College of Letters, Arts, \& Sciences, University of Southern California, Los Angeles, CA 90089, USA}
\affiliation{Center for Quantum Information Science \& Technology, University of
Southern California, Los Angeles, CA 90089, USA}
\author{Jeffrey Marshall}
\affiliation{Quantum Artificial Intelligence Laboratory (QuAIL), NASA Ames Research Center, Moffett Field, CA, 94035, USA}
\affiliation{USRA Research Institute for Advanced Computer Science (RIACS), Mountain View, CA, 94043, USA}
\author{Tameem Albash}
\affiliation{Department of Electrical and Computer Engineering, University of New Mexico, Albuquerque, New Mexico
87131, USA}
\affiliation{Department of Physics and Astronomy and Center for Quantum Information and Control, University of New
Mexico, Albuquerque, New Mexico 87131, USA}
\author{E. M. Levenson-Falk}
\email[Corresponding author: ]{elevenso@usc.edu}
\affiliation{Department of Physics \& Astronomy, Dornsife College of Letters, Arts, \& Sciences, University of Southern California, Los Angeles, CA 90089, USA}
\affiliation{Center for Quantum Information Science \& Technology, University of
Southern California, Los Angeles, CA 90089, USA}

\begin{abstract}
    Open quantum systems are a topic of intense theoretical research. The use of master equations to model a system's evolution subject to an interaction with an external environment is one of the most successful theoretical paradigms. General experimental tools to study different open system realizations have been limited, and so it is highly desirable to develop experimental tools which emulate diverse master equation dynamics and give a way to test open systems theories. In this paper we demonstrate a systematic method for engineering specific system-environment interactions and emulating master equations of a particular form using classical stochastic noise in a superconducting transmon qubit. We also demonstrate that non-Markovian noise can be used as a resource to extend the coherence of a quantum system and counteract the adversarial effects of Markovian environments.
\end{abstract}

\maketitle

 \section{Introduction}
The study of open quantum systems remains an active area of research at the frontier of understanding the range of phenomena allowed by quantum mechanics. Open systems are characterized by a system of interest having significant interactions with a number of uncontrolled environmental degrees of freedom, giving rise to decoherence in the primary system. In the case where environmental interactions take the form of purely Markovian (memoryless) decoherence, a master equation of Lindblad form (ME) \cite{Lin1976, gks} can be written and solved, in principle. However, when environmental interactions lead to non-Markovian effects, i.e. when the environment has finite-time correlations that in turn affect the system (``finite memory''), theoretical descriptions are much more challenging. The Nakajima-Zwanzig equation \cite{Zwanzig1960} provides an exact physical description of such a setup, but the equation is in general not solvable. In fact, it is difficult to even write down such an equation as it requires a complete description of the environmental degrees of freedom \cite{TOQS-book}. Simpler, more easily solved descriptions exist \cite{budini-stochastic,gen-semi-markov}, such as the post-Markovian master equation (PMME) \cite{ShabaniLidar:05}, Gaussian collapse model \cite{ferialdi_exact_2016}, quantum collisional models \cite{Budini:2013us, composite-collision, correlated-ME}, time-convolutionless master equations \cite{TOQS-book}, and the pseudo-Lindblad master equation (PLME) \cite{Groszkowski2022}. However, these are difficult to interpret physically, so it remains an open question how to write a solvable physical description of an arbitrary open quantum system.

Despite significant theoretical progress, experimental tests of open quantum system theories are more limited. Progress has been made in fitting MEs to measured dynamics \cite{Zhang2022,Groszkowski2022} and simulating Markovian environments \cite{barreiro2011open}, and techniques exist to simulate specific non-Markovian effects \cite{nm-channel-addition}, for example by embedding the system into a larger Markovian system \cite{Wang2011,m-to-nm-transition, nm-optics, oqs-zeno}. However, there is still no general experimental toolkit.
Developing new capabilities to simulate non-Markovian MEs remains highly desirable, as they would allow new experimental tests of the validity of open system models. In addition, many non-Markovian environments can be used as resources for enhancing coherence of a target system, and so this environmental engineering can be used to improve the fidelity of practical quantum processes \cite{Dong2018}. 

A particular class of non-Markovian ME, the \emph{generalized Markovian master equation} (GMME) is often exactly solvable via Laplace transforms \cite{daffer-gen-markov, marshall, engineering-gen-pauli, dynamical_beyond_markovian}. This ME describes a system undergoing Markovian dephasing while coupled to a non-Markovian environment with some finite memory. If the Lindblad operators associated with the Markovian and non-Markovian interactions act along orthogonal directions and the non-Markovian environmental memory is sufficiently long, the coherence of the system may be extended compared to the case where only the Markovian background dephasing exists. Crucially, the dynamics described by the GMME may, in some circumstances, be emulated with noisy classical driving. The GMME is thus an ideal test case for emulation of target ME dynamics with an experimental system. 

In this paper we demonstrate protocols for emulation of GMME dynamics with classical control by noisily driving a single superconducting transmon qubit. Our numerical simulations and experimental measurements conform well to the analytic solutions of the GMME in their regimes of validity. We also extend our protocol to a new regime, where the background dephasing itself is not perfectly Markovian, and model this numerically with a Bloch-Redfield master equation. We explore the limits of such regimes and describe possible extensions of this protocol. Our results provide a basis for emulation of more arbitrary open system dynamics and add another experimental tool for open system engineering.
  
\section{Background}
\subsection{Theory}
Our goal is to emulate the generalized Markovian master equation
\begin{equation}
    \frac{d}{dt} {\rho}(t) = \gamma_i \mathcal{L}_i \left(\rho(t)\right) + \mathcal{L}_j \left(\int_0^t{k(t-t')\rho(t')dt'} \right) \ ,
    \label{eq:ME}
\end{equation}
where $\mathcal{L}_i,\mathcal{L}_j$ are Lindbladians with Lindblad operators $\sigma_i,\sigma_j$ ($i,j\in\{x,y,z\}$) respectively \footnote{$\mathcal{L}_iX = \sigma_i X \sigma_i - X$.}, and $k(t-t')$ is the memory kernel of the quantum environment. This describes a system (here, a qubit) with Hamiltonian $H=0$ undergoing Markovian dephasing due to $\mathcal{L}_i$ while interacting with a non-Markovian environment via $\mathcal{L}_j$. In the case where $k(t-t') = \delta(t-t')$ the environment is fully Markovian, and the qubit state purity decays exponentially. When the environmental memory is finite, state purity decays non-monotonically and the coherence time may be extended \cite{marshall}. Note that arbitrary choices of $k$ may lead to non-physical dynamics, and so care must be taken in its selection.

To emulate this GMME we follow the recipe given in Ref.~\cite{marshall} and replace the non-Markovian environment with a stochastic classical drive given by $\hat{H}_d(t) = \frac{1}{2} B(t) \sigma_j$. We set this drive such that its classical autocorrelation function is equal to the desired quantum memory kernel, $\braket{B(t)B(t')}~=~k(t-t')$, where the expectation value is taken over many realizations of the stochastic drive. 
In order for the stochastic classical drive to emulate Eq.~\eqref{eq:ME}, the axis of the drive Hamiltonian must be orthogonal to the axis of the background Lindblad operator, i.e. $i\neq j$, so that the classical drive Hamiltonian anticommutes with the background Lindblad operator.
A derivation of how this stochastic classical drive can give rise to the GMME is given in Section~ \ref{ME_derivation}.

We choose to focus on two example memory kernels: exponentially decaying memory 
\begin{equation}
    k(t-t') = B_0^2 e^{|t-t'|/\tau_k} \ ,
    \label{eq:kernel1}
\end{equation}
and modulated decaying memory
\begin{equation}
    k(t-t') = \frac{1}{2} B_0^2 e^{|t-t'|/\tau_k} \cos(2\pi\nu (t-t')) \ .
    \label{eq:kernel2}
\end{equation}
We identify these as noise Type I and Type II respectively. For the decaying memory of noise Type I (i.e. decaying autocorrelation), we use a telegraph signal that switches between $\pm B_0$ in a Poisson process with mean switching time $\tau_k$ \cite{daffer-gen-markov}. For the modulated decaying memory of noise Type II (i.e. modulated decaying autocorrelation), we realize it with two methods. The first is to take random telegraph noise and multiply it with $\cos{(2\pi\nu t+\phi)}$, where $\phi$ is a random phase between $[0,2\pi)$. The second method utilizes the Wiener-Khinchin theorem, which states that a signal's autocorrelation is the Fourier transform of its power spectrum \cite{chatfield2003analysis}; more details are included in Section~\ref{WK_noise_generation}.

\subsection{Experimental Protocol}

Our goal is to realize the noisy drive Hamiltonians described above by subjecting our qubit to noisy control tones. We use two noise injection protocols, labeled ``XY'' and ``XZ'' after the qubit axes that the noise is injected along ($X$ being the non-Markovian component). The XZ protocol is described in detail in Section~\ref{sec:XZ}; here we describe the XY protocol. We first precisely measure the qubit transition frequency $\omega_q = \omega_{01}$ using standard Ramsey interferometry with no added noise. We perform all qubit drives at this frequency, so that in the rotating frame of the drive the qubit Hamiltonian is 0 (as required by Eq. \ref{eq:ME}) and the drive causes rotations about an axis in the $XY$ plane, with the drive phase determining the axis.

\begin {figure}[!h]
\centering
\includegraphics[width=0.85\columnwidth]{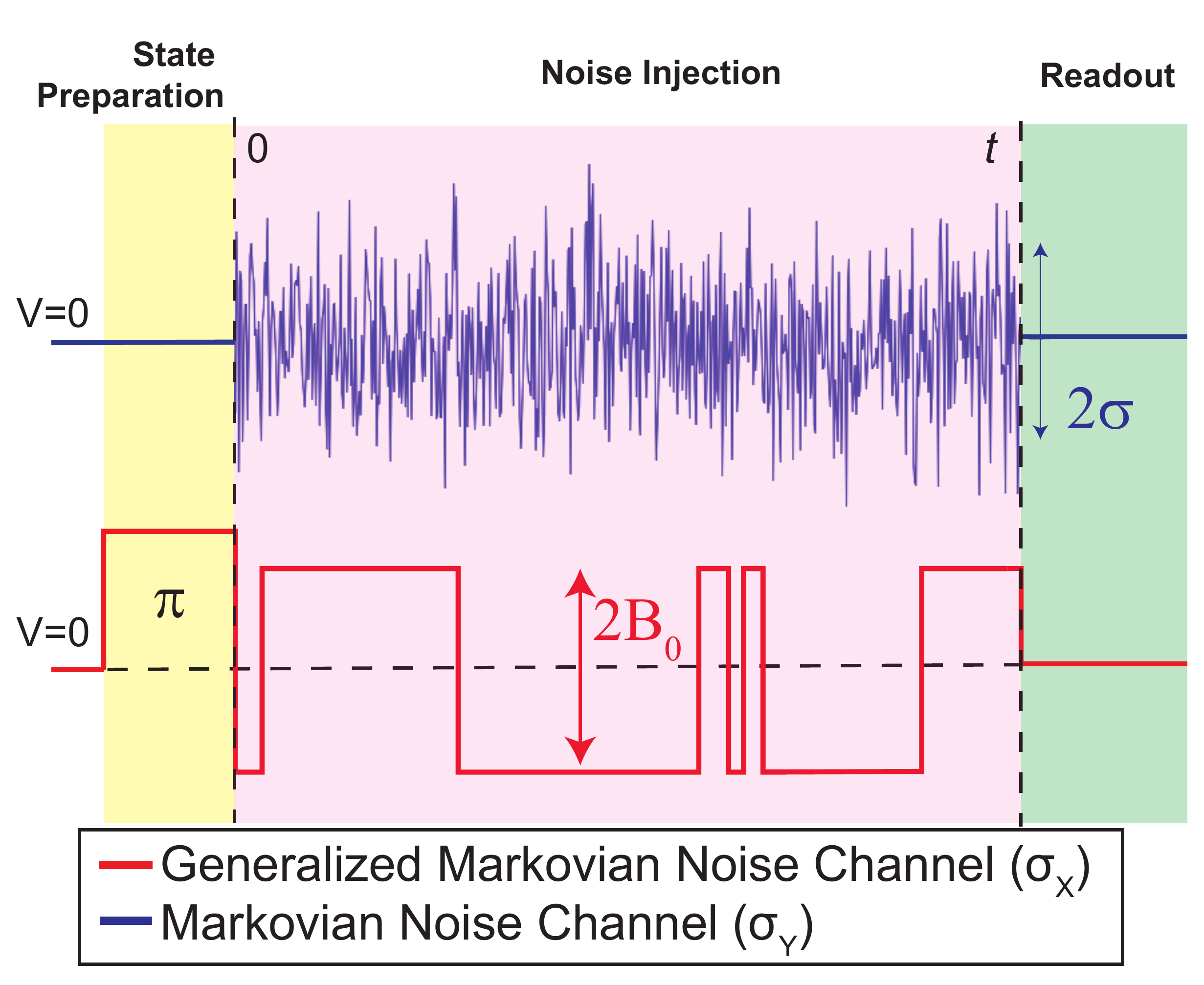}
\caption{(a) XY noise injection protocol for noise Type I. The qubit is prepared to the excited state by applying a $\pi$ pulse on the X axis (lower curve, in red). After that, white noise of variance $\sigma^2$ is injected along the Y axis via the Q channel of the IQ mixer (upper curve, in blue), while GM telegraph noise of amplitude $B_0$ is injected along the X axis via the perpendicular I channel. The amplitude of the white noise $\sigma$ is adjusted to reduce the coherence time $\tau_0$ to $\approx 1\mu$s when $B_0 = 0$. After a variable time $t$, we read out the qubit state in the Z basis. For noise Type II, the GM noise is multiplied by a cosine with random phase or is generated using the Wiener-Kinchin method. (b) Noise instances for the 3 different types of noise we inject and the corresponding memory kernels. The first waveform is a white noise instance, used to emulate a Markovian background. The second waveform is an example of random telegraph noise with $\tau_k=2\mu$s, used for noise Type I. The third waveform is generated by multiplying a random telegraph signal ($\tau_k=2\mu$s) by $cos[2\pi\nu t + \phi]$, where $\nu=2$ MHz, used for noise Type II. The total length of these waveforms is $10\mu$s.}
\label{fig:pulse_seq}
\end{figure}

We then proceed to inject noise. The pulse sequence is depicted in Figure \ref{fig:pulse_seq}. First we prepare the $\ket{1}$ state by applying a $\pi$ pulse. Noise along both $X$ and $Y$ axes ($\sigma_x$ and $\sigma_y$ terms in the drive Hamiltonian) can dephase this state, but as we see in our protocol, they can be engineered to counteract each other. We take a white noise signal sampled at 1.2 GS/s (previously generated in software) and feed it into the Q port of an IQ mixer, with a local oscillator (LO) at $\omega_q$. The output signal is a tone at $\omega_q$, $90^\circ$ phase shifted from the LO, with its amplitude modulated by the white noise signal. This is effectively a noisy stochastic $\sigma_y$ drive, which causes rapid dephasing of the $\ket{1}$ state. The goal of this is to generate a purely Markovian environment for the qubit, emulating the first term in Eq.~\eqref{eq:ME}. The result is a monotonic exponential decay in fidelity with respect to that state, characterized by a time constant $\tau_0$, which serves as a benchmark for coherence preservation later. To emulate the second term in Eq.~\eqref{eq:ME}, a stochastic signal with non-zero memory, which we refer to as generalized Markovian (GM) noise, is fed into the same mixer's I port. This ensures a phase difference of $\pi/2$ between the two drives, and so this drive is effectively a $\sigma_x$ term in the rotating frame. The effective drive Hamiltonian is $\hat{H}_d(t)=\tfrac{1}{2}\Omega_M(t)\sigma_y+\tfrac{1}{2}\Omega_N(t) \sigma_x$, where $\Omega_{M,N}(t)$ are the Markovian (white) noise and GM noise signals, respectively. After an evolution time $t$, the qubit is measured in the $\sigma_z$ basis, i.e.~with no additional pulses.  The evolution time $t$ is swept and each measurement is repeated to build up statistics and take an expectation value $\braket{\sigma_z}$ at each time point. This entire sequence is then repeated \emph{N} times, each time with a new instance of white and GM noise. The resulting \emph{N} curves are averaged over the different noise realizations and finally compared to the result of the master equation solution. We also generate simulated qubit fidelity curves under the influence of white and GM noise by numerically solving the stochastic Schr{\"o}dinger equation (SSE) and averaging over many noise realizations (i.e.~over many qubit trajectories). These simulations treat the transmon as a true qubit; we confirmed with simulations that the transmon's finite anharmonicity is not expected to have a significant effect (see Section \ref{sec:Sims} for details).
We measure and simulate the effects of GM noise over a broad range of noise parameter values, i.e.~the amplitude $B_0$, mean switching time $\tau_k$, and modulating frequency $\nu$ (for noise Type II only). Prior to each parameter point, the qubit and readout mixers are automatically calibrated to minimize leakage at $\omega_{LO}$, and the $\pi$-pulse is also re-calibrated to minimize state-preparation-related errors.

\section{Results}

\subsection{Background Markovian Dynamics}
\begin {figure}[!h]
\centering
\includegraphics[width=0.85\columnwidth]{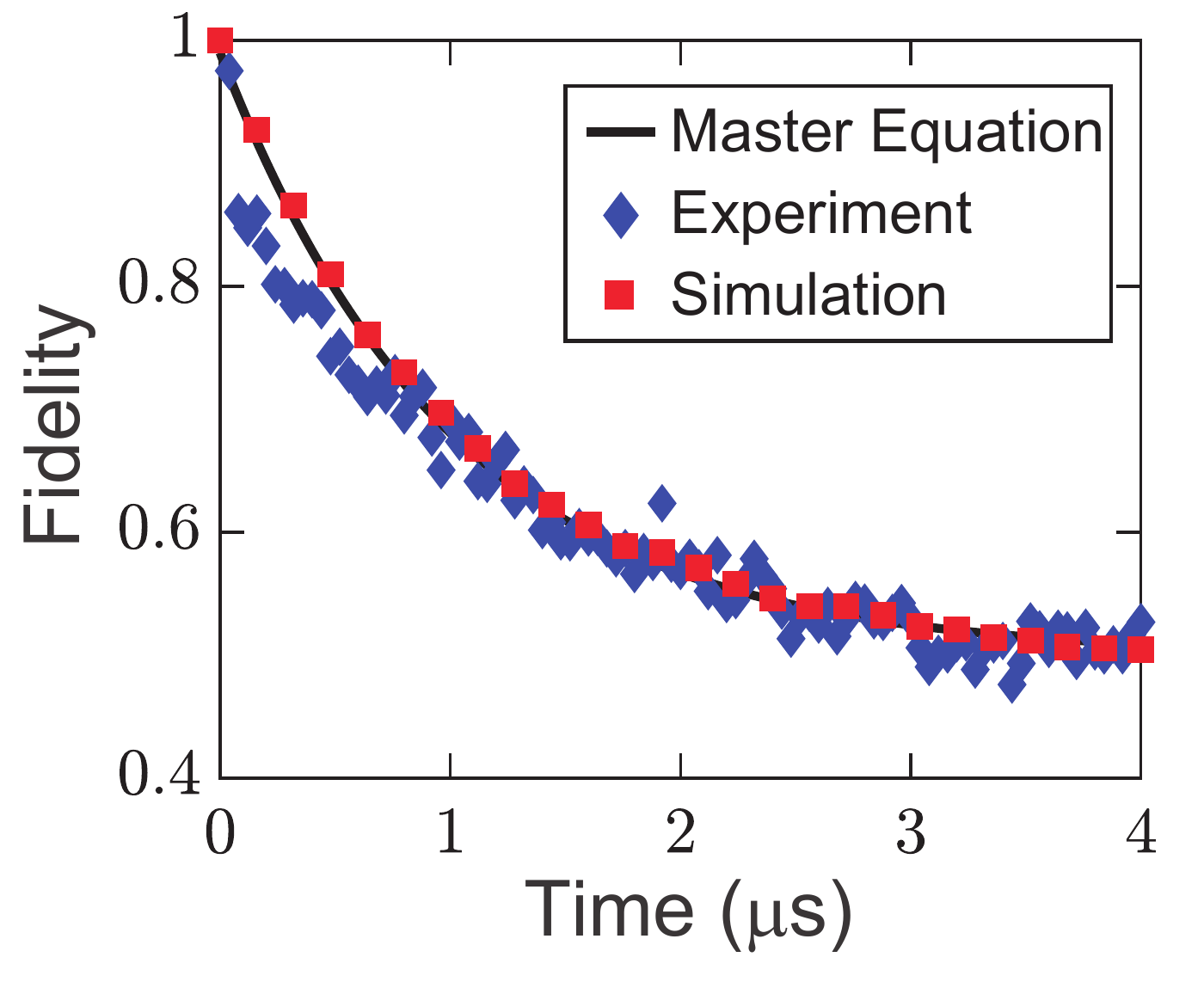}
\caption{Fidelity to the initial state under the influence of a purely Markovian background. Error bars are smaller than markers and thus omitted. Experimental data (blue diamonds) shows a monotonic exponential decay in fidelity, in good agreement with the analytic ME solution (black line) and numerical SSE simulations (red squares).}
\label{fig:bkgd_dyn}
\end{figure}

Before measuring the effect of GM noise, we first inject only white noise into the qubit in order to emulate a Markovian background. We measure state fidelity $F(\rho(t))=
\langle\psi_0|\rho(t)|\psi_0\rangle$, where $\ket{\psi_0}$ is the initial pure state, as a function of time and extract the coherence (fidelity) decay time $\tau_0$. This will later serve as a reference value for coherence enhancement. The amplitude of the white noise is adjusted to yield $\tau_0\approx1-2$ $\mu$s, down from its bare value of $\sim 100$ $\mu$s. This ensures that the dominant dephasing process is due to our injected Markovian noise. An example of qubit state fidelity under the influence of such noise is shown in Figure \ref{fig:bkgd_dyn}. The results show a monotonic, exponential decrease in fidelity as function of time. We compare experimental results with the analytic solution of the master equation and with fidelities obtained by numerically solving the SSE, averaged over simulated trajectories. The results show good agreement, indicating that the qubit is experiencing a Markovian environment to a good approximation. This measurement is repeated immediately before measuring the effects of GM noise with a given set of parameters, and the fit $\tau_0$ is used as a reference value\textemdash this accounts for any slow drifts in $\tau_0$ that may result from, e.g., mixer miscalibration.

\subsection{Noise Type I}

\begin{figure*}
\centering
 \includegraphics[width=0.8\textwidth]{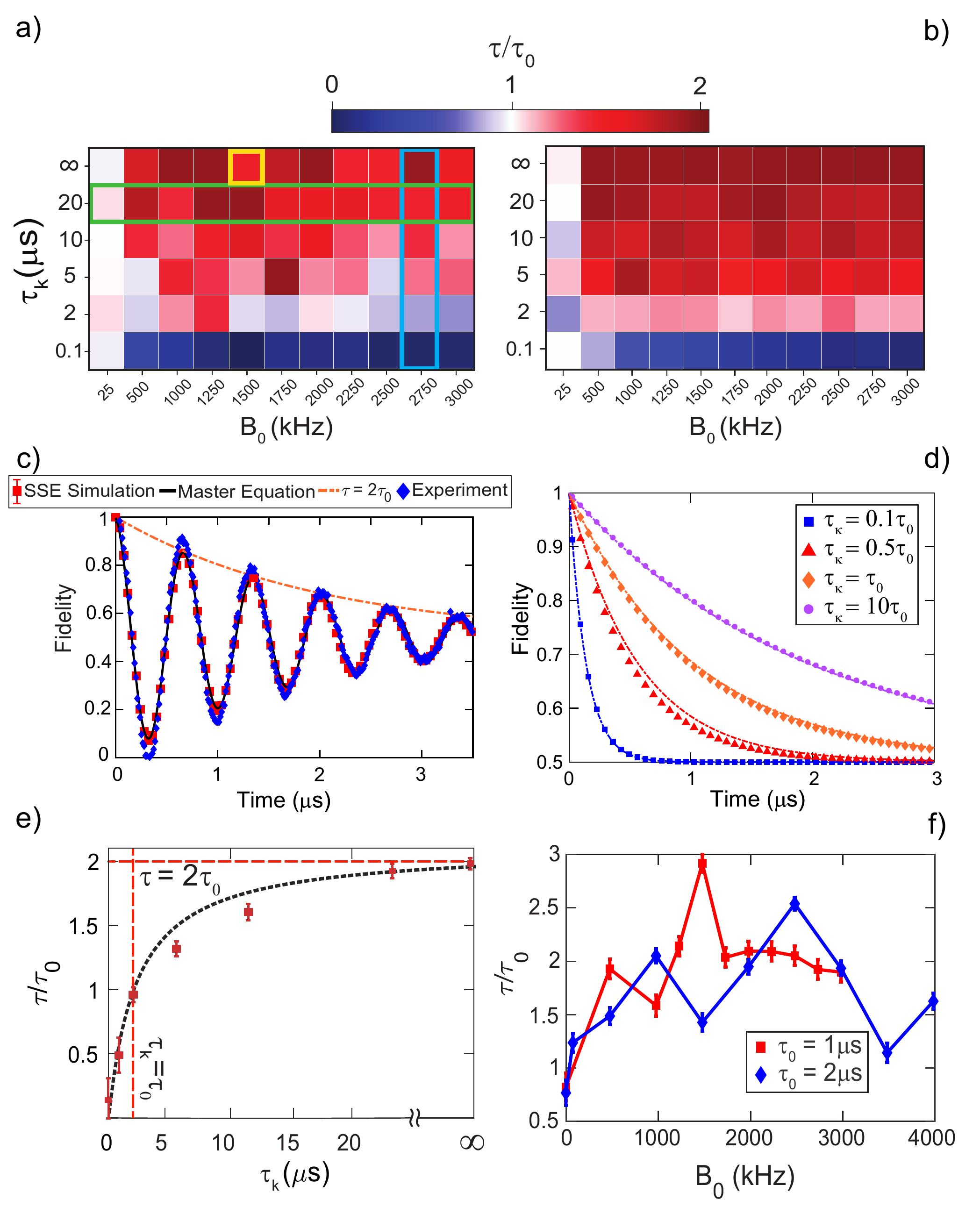}
\caption{Comparison of experimental, SSE simulation, and master equation results for noise Type I (random telegraph noise). (a,b) Enhancement of coherence, expressed as the ratio of coherence time with versus without GM noise $\tau/\tau_0$, as a function of telegraph noise amplitude $B_0$ and decay constant $\tau_k$ for (a) experiment and (b) simulation. The data and simulation show good agreement, with random variations in experimental measures of $\tau/\tau_0$ due to a finite number of noise realizations averaged and imperfect fits caused by state preparation and measurement errors. For these measurements $\tau_0\approx2\mu$s. (c) Fidelity as a function of time at a single parameter point $B_0 = $1500kHz, $1/\tau_k =0$ (highlighted with a yellow box in (a)). Error bars are smaller than markers and thus omitted. The data agrees quantitatively with the GMME solution and SSE simulation results. The envelope decays with a characteristic time of $\tau = 2\tau_0$, as expected. (d) Fidelity decay envelopes extracted from the experimental data (markers) and corresponding master equation solution (dashed lines) for different values of $\tau_k/\tau_0$ at $B_0 = 2$ MHz again showing excellent agreement.  Error bars are smaller than markers and thus omitted.(e) Coherence time $\tau$ as a function of memory decay constant $\tau_k$, and theoretical prediction, at $B_0 =$ 2750 kHz (the 1D slice shown in teal in the experimental (a)). For sufficient $B_0$, the coherence improves with increasing $\tau_k$, with the break-even point occurring when $\tau_k=\tau_0$ and a saturation at $\tau = 2\tau_0$ when $\tau_k \rightarrow \infty$. (f) Coherence enhancement $\tau/\tau_0$ as a function of the amplitude of the telegraph noise, for two different background dephasing rates. Features are repeatable run-to-run but change when the background dephasing rate is changed.}
\label{fig:NT1_plots}
\end{figure*}

Next, we measure the effect of GM noise Type I (implemented as random telegraph noise) added on top of the Markovian background. The results are shown in Figure~\ref{fig:NT1_plots}. We judge the efficacy of our emulation protocol based on two criteria: the qualitative behavior of the fidelity and the quantitative modified fidelity decay time of our qubit, which we call $\tau$. 
In Figure \ref{fig:NT1_plots}(c), we plot fidelity versus time averaged over many instances of GM and white noise for GM noise generated with strength $B_0 = 1500$ kHz and switching rate $1/\tau_k = 0$ (i.e.~infinite environmental memory). The experimental data show excellent quantitative agreement with the SSE simulations and the analytic GMME solution. The fidelity develops oscillations that decay with an exponential envelope with decay time $\tau\approx2\tau_0$. Our models assume a perfect qubit, but we have shown numerically that the analytical solution is unchanged when we include the third level. Simulation results shown in \ref{fig:qutrit_sim_plot}
We sweep the parameters of the random telegraph (GM) noise signal, i.e. amplitude $B_0$ and memory decay constant $\tau_k$, and extract fidelity decay time $\tau$ at each parameter point. Results are plotted in Figure~\ref{fig:NT1_plots}(a), given as a ratio with the measured $\tau_0$ at each point to account for small variations in $\tau_0$. We find that, as predicted in the theory, coherence is reduced for signals with small decay constant $\tau_k$ (i.e. high switching rate) and high amplitude, while the improvement saturates to twice the background value of $\tau_0$ as $B_0 \gg 1/\tau_0$ and $\tau_k\rightarrow\infty$. This is in good agreement with numerically simulated SSE results shown in Figure~\ref{fig:NT1_plots}(b). We can also extract the envelope of the fidelity decay (neglecting the oscillations) and compare this to the theoretical prediction $\tau^{-1}=(\tau_0^{-1}+\tau^{-1})/2$, as shown in Figure~\ref{fig:NT1_plots}(d) for $B_0 = 2$ MHz and several values of $\tau_k$. Again we find good agreement between experiment and theory.

We note that there is structure in the dependence of fidelity decay time on noise amplitude $B_0$, as shown in Figure~\ref{fig:NT1_plots}(f). While the ratio $\tau/\tau_0$ generally increases to a saturation value $\approx 2$ as $B_0$ increases, there are peaks and valleys in the dependence. These features are repeatable over many runs of the experiment, but the values of $B_0$ at which they appear seem to depend on the background decoherence rate $\tau_0$. At present we do not have a satisfactory explanation for this effect, but we hypothesize it may be due to some nonidealities originating from accidental transitions to higher levels of our transmon qubit.

We can gain an intuitive understanding of this prolonged coherence, and why the GM noise signal needs to act on the orthogonal axis to the background dephasing noise, using the Bloch sphere representation of the qubit. For simplicity we focus on the case of $\tau_k\rightarrow\infty$, where the stochastic GM signal becomes a constant Rabi drive with a random sign. We begin with the qubit prepared in the excited state  $\ket{1} $. Our injected white noise signal causes random rotations around the $y$ axis, which leads to dephasing. The addition of a constant $\sigma_x$ drive rotates the qubit around the $x$ axis in the $y$-$z$ plane. In the limit where $B_0 \gg 1/\tau_0$, the state completes many $x$-axis rotations before decohering. This means that, on average, the qubit state lies along the $y$ axis as much as it does along the $z$ axis. The $z$-component is affected by the white noise as before, but the component along $y$ is not, and so half the dephasing is eliminated. Thus the coherence is extended by a factor of 2. Note that if the GM noise was along the $y$ axis, this coherence preservation would not happen, as the qubit state would have $z$- and $x$-components which were both vulnerable to $y$-axis dephasing. We can also view our protocol as related to dynamical decoupling (DD), where rotations by an operator can cancel out quasi-static noise that anticommutes with that operator (in our case $\sigma_x$ and $\sigma_y$, respectively). The exact correspondence between our protocol and DD remains to be explored, as DD is typically viewed as only effective against noise which is quasi-static (i.e.~non-Markovian) on the timescale of the DD sequence \cite{Lidar_DFS_NS_DD}.

\subsection{Noise Type II}
\begin{figure*}
\centering
\includegraphics[width=0.95\textwidth]{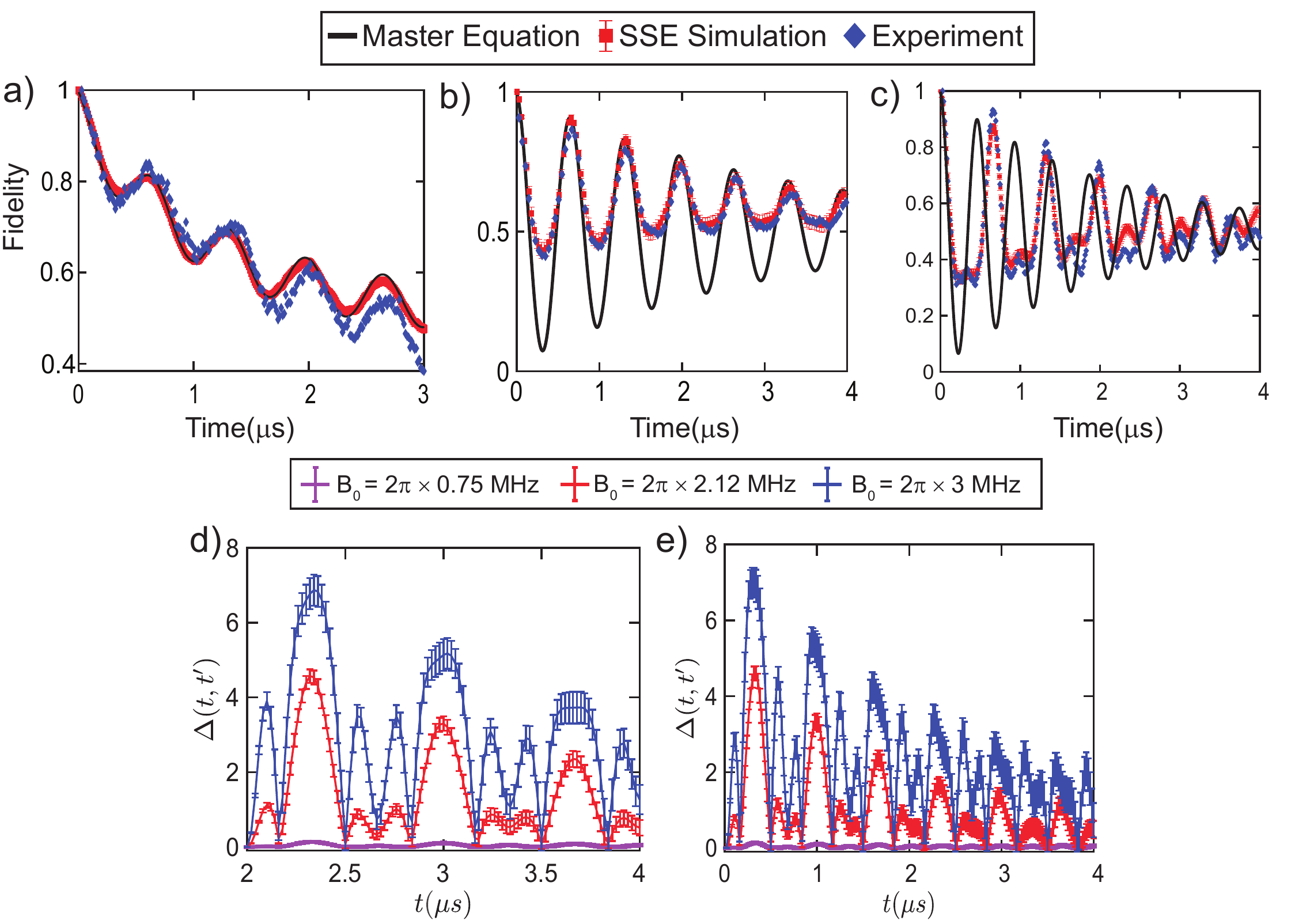}
\caption{Plots of fidelity as a function of time under the influence of noise Type II for 3 different noise amplitudes $B_0 =$ (a) 750 kHz, (b) 2122 kHz, and (c) 3000 kHz with $\nu=1500$kHz. Error bars are smaller than markers and thus omitted. The fidelity exhibits significantly different behavior based on the relationship between $B_0$ and $\nu$. We also observe that as the amplitude increases the experimental and SSE simulation results begin to diverge from the analytic GMME solution. We attribute this to the fact that the decorrelation condition necessary for ME emulation is being violated. We show the decorrelation condition $\Delta(t,'t)=\braket{B(t)B(t')\rho_{00}(t')}-\braket{B(t)B(t')}\braket{\rho_{00}(t)}$ as a function of time $t$ for (d) $t-t'=2\mu s$ and (e) $t-t'=0.02\mu s$. It is evident that when the decorrelation condition is satisfied (i.e. for the smaller amplitude $B_0=2\pi\times750$ kHz), there is good agreement between master equation and experiments.}
\label{fig:nt2_fid_plots}
\end{figure*}

We next turn to GM noise Type II\textemdash noise with a modulated decaying memory given by Eq.~\eqref{eq:kernel2}. One way to generate such a signal is to modulate random telegraph noise by multiplying the telegraph signal by $\cos(2\pi\nu t+\phi)$, where $\phi$ is a uniformly, randomly distributed phase in the $[0,2\pi)$ interval. This random phase is necessary to make the random noise process stationary.  Similar to our results for noise Type I, we find coherence is prolonged only for $\tau_k>\tau_0$. Unlike the former case though, the parameter space of amplitude $B_0$ and frequency $\nu$ exhibits more diverse features. The fidelity no longer exhibits decaying oscillations of the form $\cos(\omega t)e^{-t/\tau_k}$ for all combination of noise parameters. Results for three different combinations of amplitude $B_0$ and frequency $\nu$ of the GM noise signal are shown in Figure~\ref{fig:nt2_fid_plots}. We can group the fidelity results into 3 empirical groups, based on the relationship between $B_0$ and $\nu$: 
\begin{description}
    \item [$B_0\ll\nu$] The fidelity exhibits oscillations of the form $A\cos(\omega t)+B\exp{(-t/\tau)}+C$. 
    \item [$B_0 \approx \sqrt{\frac{2}{9}(\frac{1}{\tau_0}-\frac{1}{\tau_k})^2+(2\omega)^2}$, $\omega=2\pi\nu$]   
    The fidelity has the form: $A\cos(\omega t)\exp{(-t/\tau)}+C$
    \item [$B_0\gg\nu$] The fidelity has the form $A_1\cos(\omega_1 t)e^{-t/\tau_1}+A_2\cos(\omega_2 t)e^{-t/\tau_2} + C$. 
\end{description}
The relationship between $B_0$ and $\nu$ in the second case was chosen because it makes the master equation analytically solvable \cite{marshall}. There is excellent agreement among master equation solution, simulation, and experiment for small amplitude $B_0$, as we can see in Figure \ref{fig:nt2_fid_plots}(a). As the amplitude increases, the envelope of the experimental and simulated fidelities match the analytic result, but the quantitative behavior does not. We attribute this discrepancy to a breakdown of the assumptions needed for ME emulation. Specifically, we define the decorrelation condition for a GM noise instance given by $B(t)$:
\begin{equation} \label{eqt:DecorrelationCondition}
\braket{B(t)B(t')\rho(t')}\approx\braket{B(t)B(t')}\braket{\rho(t)} \ .
\end{equation}
This is a a necessary assumption when deriving Eq.~\ref{eq:ME} from a classical stochastic drive \cite{marshall}. As the amplitude $B_0$ grows, this condition begins to break down. Figure \ref{fig:nt2_fid_plots}(d) shows that the larger the noise amplitude, the larger the correlation between the noise signal and the qubit dynamics. Simulation and experimental results continue to show good agreement for higher amplitudes, indicating that our results are due to a breakdown of ME emulation and not to experimental nonidealities. These results show the limits of our ME emulation technique. 

\begin{figure*}
\centering
\includegraphics[width=0.9\textwidth]{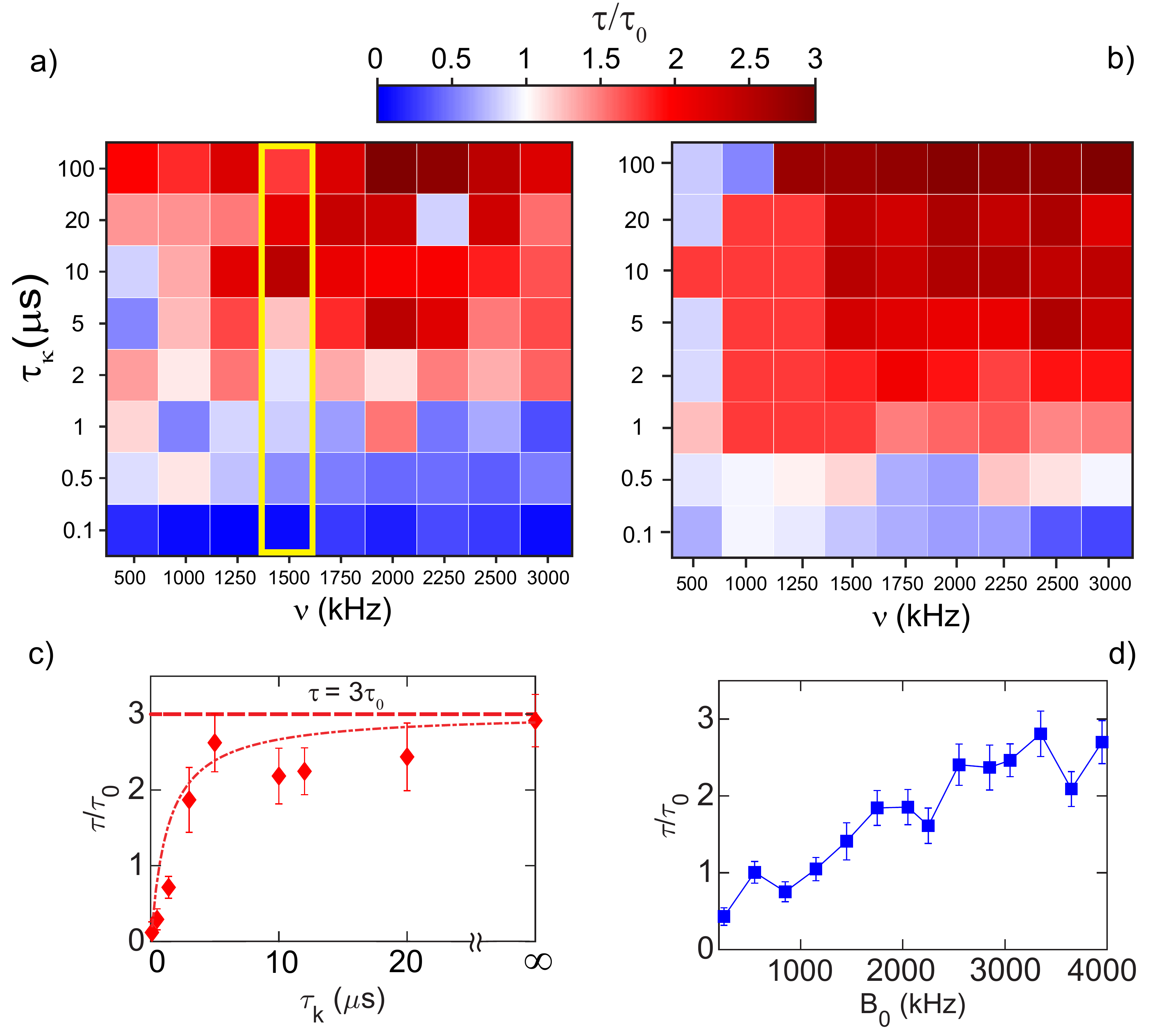}
\caption{Fidelity under the influence of noise Type II signals (modulated decaying memory kernel) implemented as modulated telegraph noise. Experimental (a) and simulated (b) coherence enhancement as a function of noise modulation frequency $\nu$ and memory time $\tau_k$. For these measurements $\tau_0\approx1.2\mu$s.  As with noise Type I, coherence enhancement increases with higher $\nu$,$\tau_k$, this time asymptoting at a higher ratio $\tau/\tau_0 \approx 3$. (c) 1D slice of the data shown in (a) (highlighted in a yellow box in (a)) showing coherence improvement as a function of memory time $\tau_k$ for $\nu=1500$ kHz, $B_0=2122$ kHz. Coherence is prolonged with longer environmental memory, asymptoting at $\tau = 3 \tau_0$. The dashed curve is a theoretical prediction from Ref.~\cite{marshall}. (d) Coherence enhancement as a function of GM noise amplitude $B_0$ for $\nu=1500$ kHz and $1/\tau_k =0$. The improvement approaches but does not exceed the theoretical limit of $3\tau_0$ as $B_0$ increases.}
\label{fig:sweeps_nt2}
\end{figure*}

We explore the noise Type II parameter space by sweeping $\nu$ and $\tau_k$. We choose $B_0$ such that the resulting fidelity has the form $A\cos(\omega t)e^{-t/\tau}+C$. The results are shown in Figure \ref{fig:sweeps_nt2}. There is qualitative agreement between simulation and experimental data. In both cases, the fidelity decay time increases as $\nu$ and $\tau_k$ increase.  The 1D slices of the 2D data shown in Figure \ref{fig:sweeps_nt2}(c)-(d) show that coherence cannot be improved beyond the $3\tau_0$ limit, as predicted by the theory. The analytic solution of the GMME predicts $\tau^{-1} = (\tau_0^{-1}+2\tau_k^{-1})/3$, as shown by the dashed curve in  Figure \ref{fig:sweeps_nt2}(c), showing good quantitative agreement of the experimental and predicted coherence times.

\begin{figure}
\centering
\includegraphics[width=0.85\columnwidth]{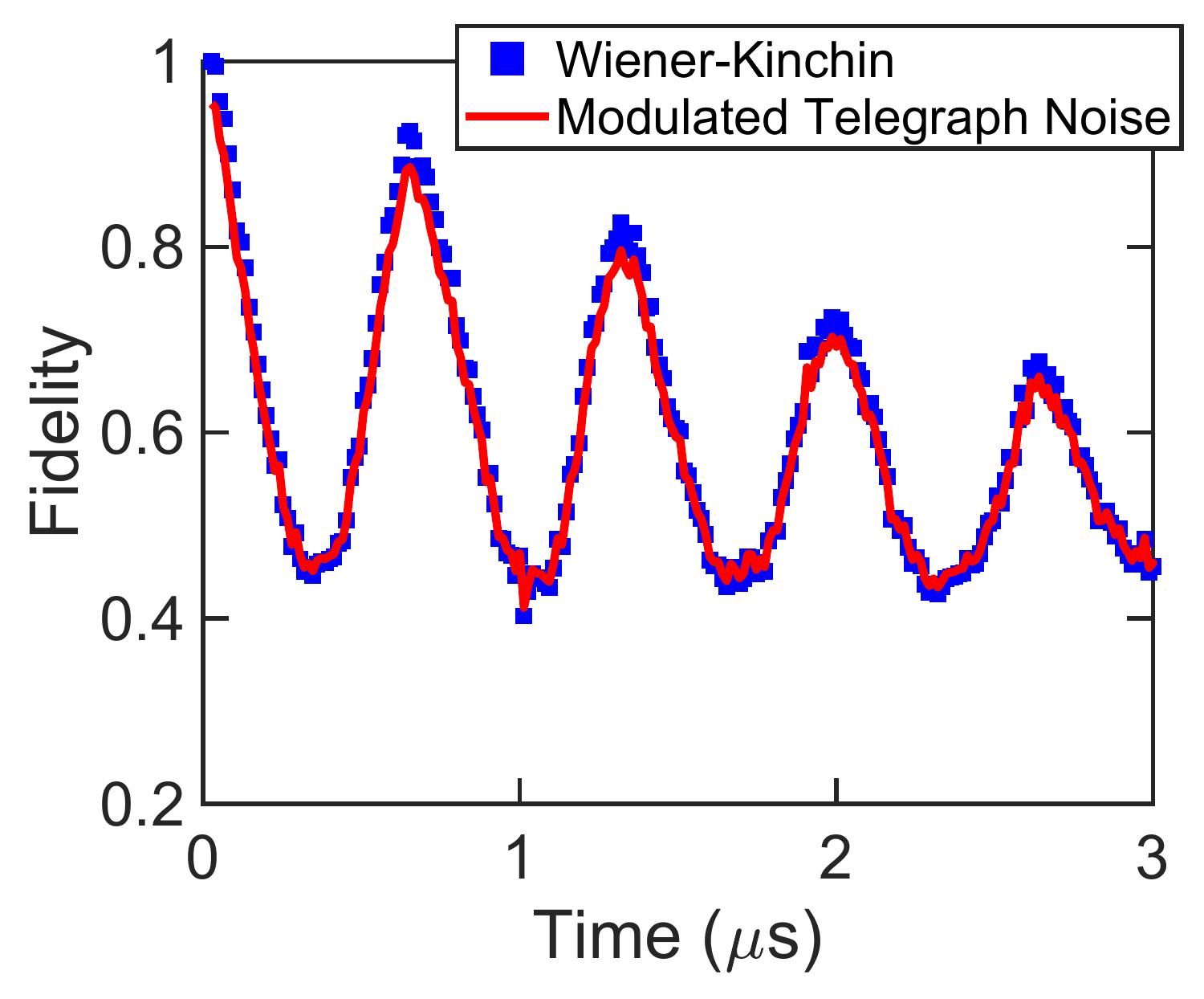}
\caption{Comparison of experimental fidelity data using signals generated by applying a periodic modulation to the telegraph noise and those generated using the Wiener-Kinchin method. These curves were generated with $B_0=2122$ kHz, $\nu=1500$ kHz, and switching rate $1/\tau_k=0$; similar agreement is found for all parameters tested.  Error bars are smaller than markers and thus omitted.}
\label{fig:WK_nt2}
\end{figure}

We also injected signals generated using the Wiener-Kinchin method described in \ref{WK_noise_generation}. This protocol is extremely flexible, as it allows us to easily generate noise waveforms from any memory kernel, without restricting ourselves to noise that has an easy analytic expression for its time series (such as the modulated telegraph noise). We tested this protocol for noise Type II with both experimental measurements and SSE simulations. We found excellent agreement between data/simulation with Wiener-Kinchin noise versus modulated telegraph noise. Results are shown in Figure \ref{fig:WK_nt2}.

\subsection{Effect of limited qubit bandwidth on protocol}\label{sec:XZ}
\begin{figure}[!h]
    \centering
    \includegraphics[width=0.85\columnwidth]{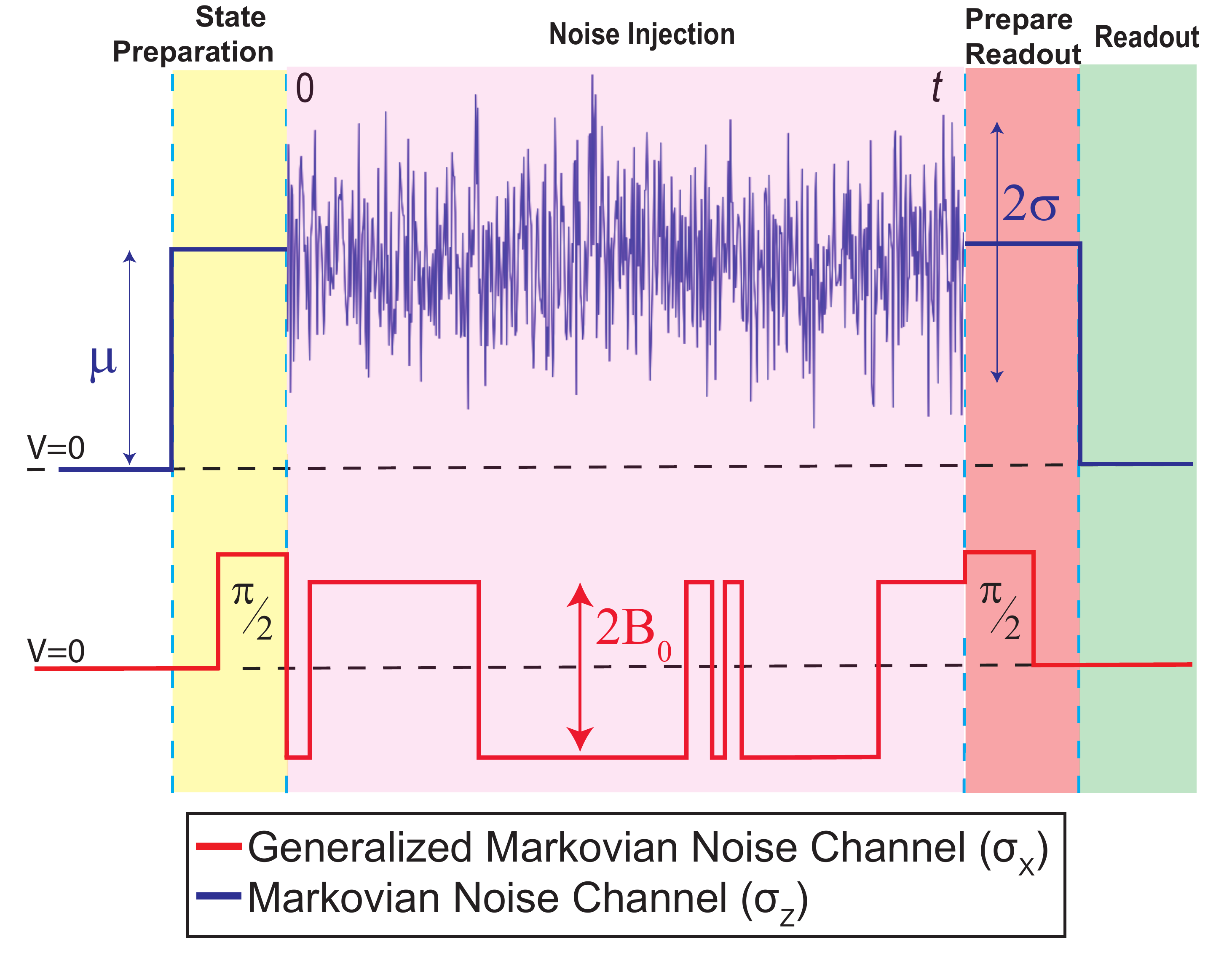}
    \caption{Control sequence for XZ protocol. The qubit frequency is first shifted by a constant AC Stark tone with an amplitude $\mu$ (upper curve in blue). The qubit is then prepared in the $\ket{+i}$ state with a $\pi/2$ pulse. Broadband white noise with variance $\sigma^2$ is injected on the AC stark tone, while GM noise (shown schematically as a telegraph signal) is injected on the qubit drive. The value of $\mu$ is adjusted such that the qubit drive pulses are on resonance with the qubit. The readout cavity heavily filters the AC Stark tone, leading to a finite-bandwidth $z$-axis noise reaching the qubit. After some evolution time $t$, the qubit is projected to the $z$-axis with another $\pi/2$ pulse and is read out.}
    \label{fig:XZ_sequence}
\end{figure}

The GMME was derived assuming a perfectly Markovian background, which we emulate by injecting white noise. However, often qubits experience strongly non-Markovian backgrounds, due to effects such as 1/f noise \cite{Burkard2009} and quantum crosstalk \cite{Tripathi2022}. Of particular interest are environments that are \emph{approximately} Markovian over long enough time scales, but have finite bandwidth, and so are non-Markovian over short times \cite{mortezapour2017non}. At some level all systems must behave this way, as no physical process is instantaneous, so an understanding of such environments is extremely desirable. To emulate such an environment, we inject heavily filtered noise that has a white spectrum within a certain bandwidth and then falls off rapidly at higher frequencies. We use the ``XZ" protocol to achieve this effect. We test this protocol with the initial state $\ket{+i}=(\ket{0} + i \ket{1})/\sqrt{2}$, prepared with a $\pi/2$ qubit rotation. Background noise is still generated as a white noise waveform, but we now mix this waveform with a carrier tone at $\omega_{Stark}$, 100 MHz detuned from the qubit readout cavity. The narrowband cavity heavily filters the noise, and the noise that reaches the qubit modifies its frequency (i.e. causes $z$-axis rotations) via the AC Stark effect \cite{Schuster2005}. After a period of time, the qubit is projected back to the $z$-axis by another $\pi/2$ pulse. The pulse sequence is shown in Figure \ref{fig:XZ_sequence}.

\begin{figure}[!h]
    \includegraphics[width=0.8\columnwidth]{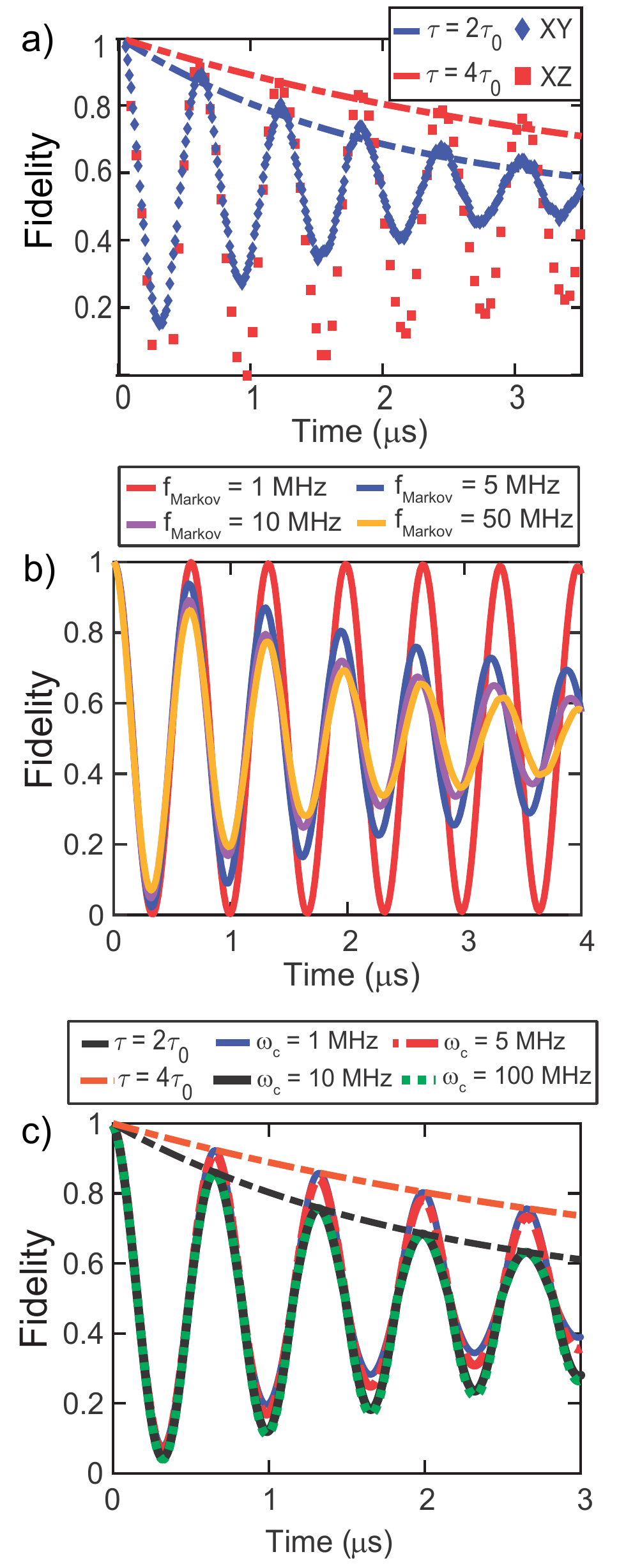}
    \caption{Effect of white noise bandwidth on the enhancement of coherence from added GM noise. (a) Experimental results from the XY and XZ protocols, for $B_0$ = 1623 kHz, $\tau_k$ = 20$\mu s$. The XY (broadband) protocol gives the theoretically-predicted coherence time enhancement of a factor of 2, while the XZ (finite bandwidth) protocol gives a factor of 4. (b) SSE simulation results of finite-bandwidth noise, generated as white noise with a variable low sampling frequency of $f_{\mathrm{Markov}}$ (compared to the XY frequency of 1.2 GS/s). (c) Bloch-Redfield simulation results for background dephasing noise with a flat spectrum with high-frequency cutoff $\omega_c$.}
    \label{fig:BW_plots}
\end{figure}

When injecting GM noise Type I on top of this finite-bandwidth background, the resulting coherence improvement was more than the theoretical limit of $2\tau_0$, up to $10\tau_0$ in some cases of high amplitude $B_0$. Experimental results comparing the XY and XZ protocols are shown in Figure \ref{fig:BW_plots}(a). We confirm that this excess improvement is due to white noise being low-pass filtered by the cavity using quantum trajectory SSE simulations (Figure \ref{fig:BW_plots}(b)), and Bloch-Redfield master equation (BRME) simulations (Figure \ref{fig:BW_plots}(c)). In the SSE simulations, we simulate the effect of a finite cutoff frequency by using white noise with a low sampling frequency, while in the BRME simulations, we have a flat frequency spectrum for our background noise with a finite frequency cutoff $\omega_c$. Both sets of simulations show that a non-white background noise will cause $\tau$ to exceed $2\tau_0$ as $\tau_k\rightarrow\infty$. The lower the frequency cutoff, the greater the increase in $\tau$. Future studies may further explore the correspondence between finite-frequency noise and non-Markovian ME dynamics.

\section{Conclusion}
We have demonstrated the emulation of the solveable GMME using classical noise with both numerical simulations and experimental tests. We find that for a purely decaying environmental memory kernel we are able to near-perfectly emulate the ME with any choice of parameters, while for a modulated decaying memory the emulation is only successful in the regime where the emulated non-Markovian noise is relatively weak. The reason for this limited success is due to the breakdown of the decorrelation condition in Eq.~\eqref{eqt:DecorrelationCondition} needed to derive the GMME from the SSE. We have also extended the GM noise injection protocol to the case where the background dephasing has a finite bandwidth and found qualitative agreement between our results and a Bloch-Redfield master equation treatment.

Our results demonstrate the utility and the limits of emulating quantum non-Markovian environments with noisy classical drives. Future work may build on these results, combining such noisy drives with engineered dissipation \cite{harrington2022engineered}, tunable couplings \cite{mostame2017emulation}, and customized baths with many degrees of freedom. It will also be interesting to use such techniques to further explore the correspondence between master equations and dynamical decoupling techniques that treat the environment as a classical noise source \cite{Biercuk_2011}. An open systems ME-emulation approach provides a new way to analyze dynamical decoupling, especially in the presence of non-ideal pulses. Our results open a new avenue in open quantum systems experimentation.

\section{Acknowledgements}
The authors thank P.~Zanardi and L.~Campos Venuti for useful discussions. Work at USC (EV, HZ, EMLF) was supported by NSF OMA-1936388, ONR N00014-21-1-2688, Cottrell Scholars Program 27550, and ARO W911NF-19-1-0070. Work at UNM (TA) was supported by NSF OMA-1936388. JM is grateful for support from NASA Academic Mission Services, Contract No. NNA16BD14C. 
JM's contribution to this work was additionally supported by the U.S. Department of Energy, Office of Science, National Quantum Information Science Research Centers, Superconducting Quantum Materials and Systems Center (SQMS) under the contract No.~DE-AC02-07CH11359 through NASA-DOE interagency agreement SAA2-403602.
Qubits were provided by MIT Lincoln Laboratory via the SQUILL Foundry. Parametric amplifiers were provided by NIST Boulder. 

\section{Methods}

\subsection{Derivation of Generalized Markovian Master Equation} \label{ME_derivation}
In our derivation of Eq.~\eqref{eq:ME}, we follow closely the derivation in Ref.~\cite{marshall}. We begin by adding a stochastic Hamiltonian $H(t)=\frac{1}{{2}}B(t) \sigma_j$, where $B(t)$ is the generalized Markovian noise signal, on top of the background Markovian dynamics characterized by the Lindbladian $\mathcal{L}_{i}$, where $\mathcal{L}_i \rho=  \left( \sigma_i \rho \sigma_i - \rho \right)$.
Assuming we are on resonance with the qubit, i.e. $H_q=0$, the time-evolution dynamics are then described by 
\begin{equation}
    \frac{d}{dt}{\rho}(t) = \gamma_i \mathcal{L}_i\rho(t)-i[H(t),\rho(t)] \tag{M1}  \ . \label{eq:M1}
\end{equation}
A formal solution to this equation is given by:
\begin{equation}
    \rho(t) = \rho(0) + \gamma_i \mathcal{L}_i\int_{0}^{t} \rho(t')dt' -i\int_{0}^{t} [H(t'),\rho(t')]dt' \tag{M2} \ .\label{eq:M2}
\end{equation}
We then insert Eq.~\eqref{eq:M2} into the right hand side of Eq.~\eqref{eq:M1}, yielding
\begin{equation}
    \begin{split}
    \frac{d}{dt} \rho(t) &= \gamma_i \mathcal{L}_i\rho(t) - i[H(t),\rho(0)] \\
    & - i \frac{\gamma_i}{{2}}\mathcal{L}_i \int_{0}^{t}  B(t)[\sigma_j,\rho(t')dt'] \\
    &- \frac{1}{4}\int_{0}^{t} B(t)B(t')[\sigma_j,[\sigma_j,\rho(t')]dt' \ .    
   \end{split}
   \tag{M3}\label{eq:M3}
\end{equation}
In order to arrive at Eq.~\eqref{eq:ME}, we assume that  the GM signal obeys the following statistics: 
\begin{enumerate}
    \item $\braket{B(t)} = 0$\ ;
    \item $\braket{B(t)B(t')} = k(t-t')$ \ ;
    \item $\braket{B(t)B(t')\rho(t')} \approx \braket{B(t)B(t')}\braket{\rho(t')}=k(t-t')\braket{\rho(t')}$
    \item $\langle B(t)\rho(t')\rangle \approx \langle B(t)\rangle \langle \rho(t')\rangle = 0$
\end{enumerate}
where the averaging is done over the noise realizations. The second requirement means that the stochastic process is stationary, i.e. the mean and variance do not change with time. The third and fourth requirements translate to the state being sufficiently decorrelated from the stochastic noise process.  With these assumptions, when we take the average over noise realizations we arrive at
\begin{equation}
    \frac{d}{dt}{\rho}(t) = \gamma_i \mathcal{L}_i\rho(t) + \frac{1}{2} \mathcal{L}_j\int_0^t{k(t-t')\rho(t')dt'} \ ,
    \tag{M4}\label{eq:M4}
\end{equation}
which is the GMME. We solve Eq.~\eqref{eq:M4} using the Laplace transform method. This yields 
\begin{equation}
      \left( s-\gamma_i \mathcal{L}_i-\frac{1}{2} \tilde k(s)\mathcal{L}_j \right) \tilde \rho(s) = \rho(0) \ , \tag{M5}\label{eq:M5}
\end{equation}
where $\tilde{k}(s)$  is the Laplace transform of the memory kernel. The initial state can be written as $\rho_0 = 1/2 (\mathbb{I}+\overrightarrow \lambda \cdot \overrightarrow \sigma)$ and for noise type I the memory kernel transform is given by $\tilde{k}(s)=\tfrac{B_0^2}{s+1/\tau_k}$. To simplify our calculations, we are going to use the damping basis because the Pauli matrices are the eigenstates of $\mathcal{L}_i$, and hence we can replace the operators with the corresponding eigenvalues, i.e. $\mathcal{L}_y \sigma_x = -2 \sigma_x$. 

In the case of the XY protocol we have $i = y$ and $j = x$. 
Assuming $\rho_0=\ket{1}\bra{1}$, we have $\lambda_x=\lambda_y=0$ and $\lambda_z=-1$. Solving Eq.~\eqref{eq:M5} we get 
\begin{equation}
\begin{split}
   \tilde \rho(s) &= \frac{1}{2} \left(\frac{1}{s} \mathbb{I} -
     \frac{1}{s+2\gamma_y + \tilde{k}(s)
}\sigma_z \right) \ .
    \end{split}
    \tag{M6}\label{eq:M8}
\end{equation}
We note that in the case of no GM noise ($\tilde{k}(s)=0$), the fidelity decays as $e^{-2\gamma_yt}$, with $2\gamma_y=1/\tau_0$. With added GM noise, the coherence time is modified, and in the limit of $\tau_k\rightarrow\infty$, the decoherence rate is half of the original value, i.e. $\tau=2\tau_0$.  
\subsection{Noise Generation using the Wiener-Kinchin Theorem} \label{WK_noise_generation}

Noise waveforms of arbitrary memory kernels can be generated using the Wiener-Kinchin theorem, which states that the power spectral density (PSD) of a signal is the inverse fourier transform of its autocorrelation. For a stochastic signal whose timeseries is described by $B(t)$, we have
\begin{equation}
    \braket{B(t)B(t-\tau)}=\tfrac{1}{2\pi}\int_{-\infty}^{\infty} e^{i\omega t}|\tilde{B}(\omega)|^2 d\omega \ ,
\end{equation}
where $\tilde{B}(\omega)$ is the fourier transform of $B(t)$. Knowing the PSD of the signal, we can construct the fourier spectrum of the signal and the timeseries $B(t)$ by applying an inverse fourier transform. Hence, our procedure for generating noise signals with this method is the following:
\begin{enumerate}
    \item Apply the discrete inverse fourier transform to the memory kernel to obtain the PSD 
    \begin{equation}
        \tfrac{P(\omega)}{d\omega}=|\tilde{B}(\omega)|^2 \ .
    \end{equation}
    \item Multiply the result by the bin size $d\omega=1/T_{max}$, where $T_{max}$ is the length of the final timeseries, to get the power at each frequency.
    \item Construct the fourier spectrum by multiplying the power at each frequency with a random phase factor $e^{i\phi}$, where $\phi\in[0,2\pi)$.
    \item Apply the inverse fourier transform to obtain the timeseries $B(t)$.
\end{enumerate}

\subsection{Noise Injection} \label{protocol_description}
All noise waveforms and qubit manipulation pulses for our experiment are generated digitally at room temperature using an Arbitrary Waveform Generator (HDAWG) from Zurich Instruments at 1.2 GS/s sampling rate. XY noise waveforms are upconverted from DC to microwave frequencies using an IQ mixers with an LO at $\omega_q/2\pi = 3.3321$ GHz, where $\omega_q$ is the qubit 01 frequency. Z noise waveforms are upconverted to $\omega_{Stark}/2\pi= \omega_r/2\pi + 100$ MHz$=7.3583$ GHz. The readout signal is generated with a sampling rate of 1.8 GS/s using a Quantum Analyzer (UHFQA) from Zurich Instruments, and is upconverted to $\omega_r/2\pi= 7.2583$ GHz before injection into the fridge. After transmitting through the qubit's measurement cavity, the readout tone is amplified by a Josephson Parametric Amplifier (JPA) at base temperature, a semiconductor amplifier at 3.5 K, and semiconductor amplifiers at room temperature. The JPA is pumped with a flux tone at $\omega_p=2\pi\times14.53$ GHz $\approx 2\omega_r$, giving $\sim 20$ dB gain at $\omega_r$. The amplified readout signal is demodulated and digitized at 1.8 GS/s in the same Quantum Analyzer, where it is integrated and stored as a single numeric value. This value is subjected to a threshold test to determine the qubit state; in the case that the qubit state is $\ket{1}$, a pulse is sent back to the HDAWG that triggers it to send a $\pi$ pulse to the qubit, thus deterministically resetting the qubit to the ground state. This greatly speeds the experimental parameter sweep, as the reset process takes only $\sim 10$ $\mu$s, much faster than waiting for the qubit to decay naturally ($T_1 \sim 100$ $\mu$s). Qubit experimental parameters are listed in Table \ref{tab:qubit_char_table}. 

\begin{figure}
    \centering
    \includegraphics[width=0.85\columnwidth]{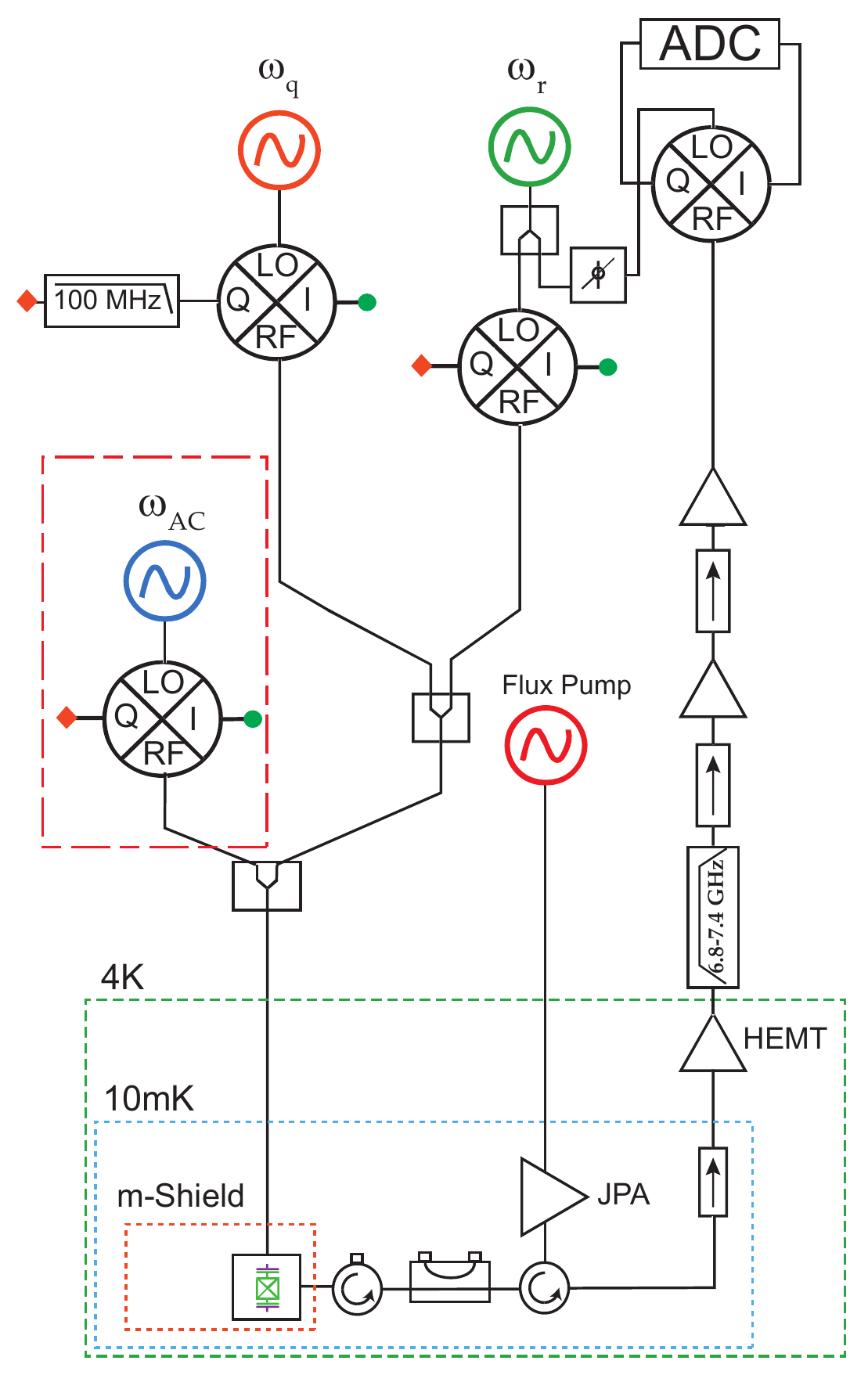}
    \caption{Experimental setup. The orange diamonds and green circles correspond to AWG channels which are used to generate state preparation pulses, noise waveforms, and readout pulses. The 100 MHz low-pass filter between the AWG and the qubit mixer's Q channel is used to prevent white noise waveforms in the XY protocol from accidentally driving the $\ket{1}\rightarrow\ket{2}$ transition. The mixer enclosed in red is only used in the XZ protocol.}
    \label{fig:fridge_setup}
\end{figure}

All instruments are integrated and controlled via a Python API. The workflow process for a single point in the noise parameter space is as follows,
\begin{enumerate}
    \item \emph{N} GM noise realizations with non-zero memory are generated by a Python script and stored into an array.
    \item \emph{N} Ramsey measurements are performed with only white noise injected into the qubit for benchmarking of the background dephasing rate $\tau_0$.
    \item \emph{N} Ramsey measurement are performed with both white and GM noise. These measurements are interleaved with the white-noise-only measurements.
    \item Ramsey traces are averaged over the noise realizations and fitted to the appropriate function to extract the parameters of interest, $\tau_0$ and $\tau$.
\end{enumerate}
The detailed experimental setup is shown in Figure \ref{fig:fridge_setup}.

\begin{table*}[h]
\centering
\begin{tabular}{c  c  c  c  c  c  c  c c c}
$\omega_{01}/2\pi$ (GHz) & $\omega_{Stark}/2\pi$ (GHz) & $\omega_r/2\pi$ (GHz) & $2\chi/2\pi$ (kHz) &  $\alpha/2\pi$ (MHz) & $\kappa/2\pi$ (kHz) & $g/2\pi$ (MHz) & $T_1 (\mu s)$ & $T_2^R (\mu s)$ & $T_2^E (\mu s)$  \\
\hline\hline
3.3321 & 7.3586 & 7.2586 & 220 & -172 & 220 & 64 & 98 & 42 & 60 \\
\end{tabular}
\caption{Qubit characteristics table. From left to right: the qubit 01 transition frequency $\omega_{01}$, AC-stark frequency $\omega_{Stark}$,cavity resonant frequency $\omega_r$, cavity dispersive shift $2\chi$, cavity linewidth $\kappa$, qubit-cavity coupling strength $g$, qubit relaxation time $T_1$, qubit pure dephasing time $T_2^R$, and qubit dephasing time $T_2^E$.}
\label{tab:qubit_char_table}
\end{table*}

\subsection{Noise Parameter Calibrations}
\subsubsection{$B_0$}
\begin{figure}
    \centering
    \includegraphics[width=0.85\columnwidth]{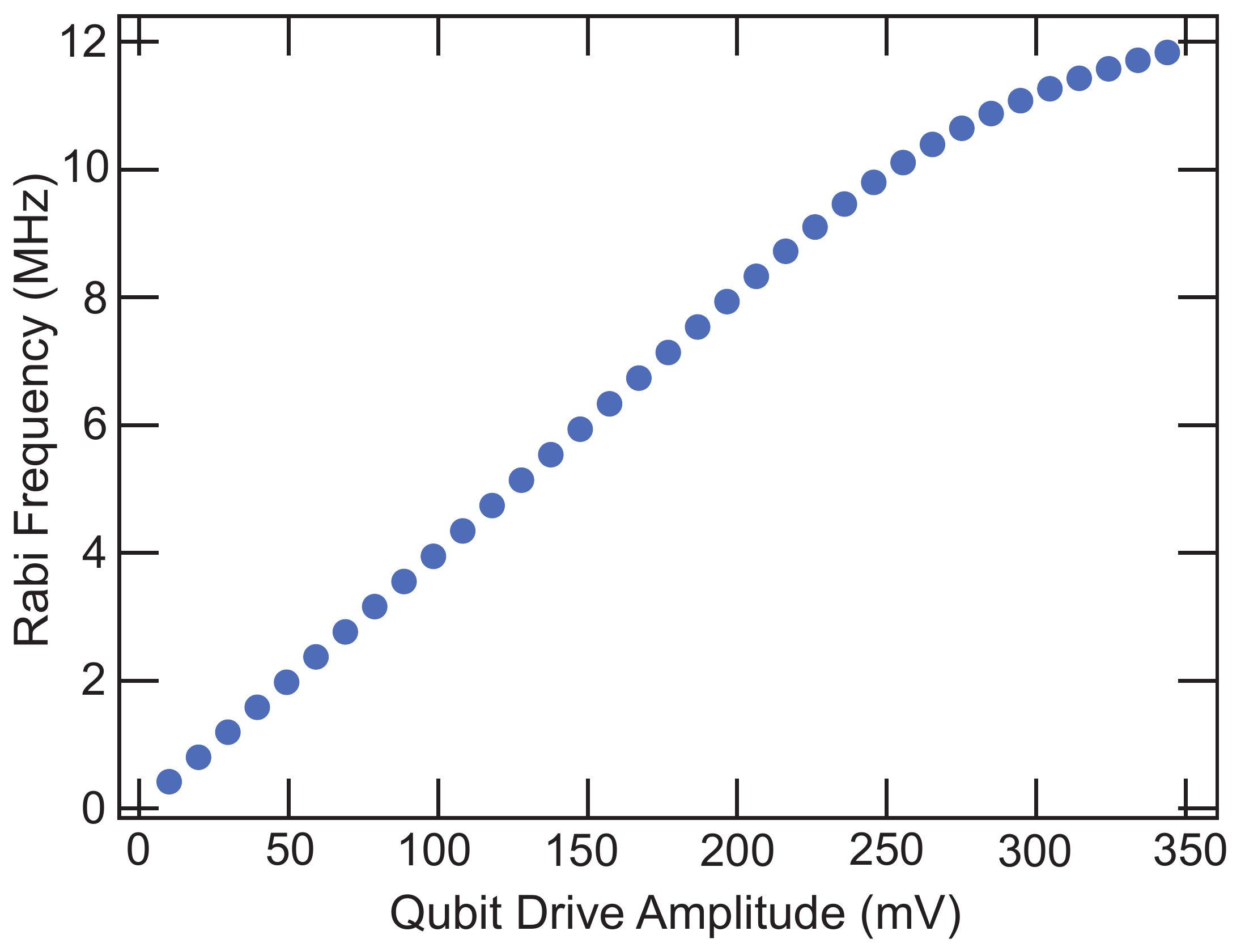}
    \caption{Rabi frequency as a function of qubit drive amplitude (X-channel). For our experiments we stay in the linear regime. The non-linear behavior for amplitudes greater than 275 mV is due to mixer saturation.}
    \label{fig:B0_cal}
\end{figure}

In order to be able to compare our experimental results with those of the simulations and the master equation solution we need to translate the characteristic noise parameters in frequency units. For the calibration of the noise amplitude $B_0$ of the generalized Markovian signal we performed time-Rabi measurements, for a range of qubit drive amplitudes. This is due to the fact that in the intervals where the random telegraph noise is not switching between high and low states it is essentially a constant Rabi drive. Thus, the noise applied at the qubit frequency can be treated as a constant Rabi drive with switching polarity. The results of the calibration are shown in Figure \ref{fig:B0_cal}.
\subsubsection{AC Stark Noise Parameters}
\begin{figure}
    \centering
    \includegraphics[width=0.85\columnwidth]{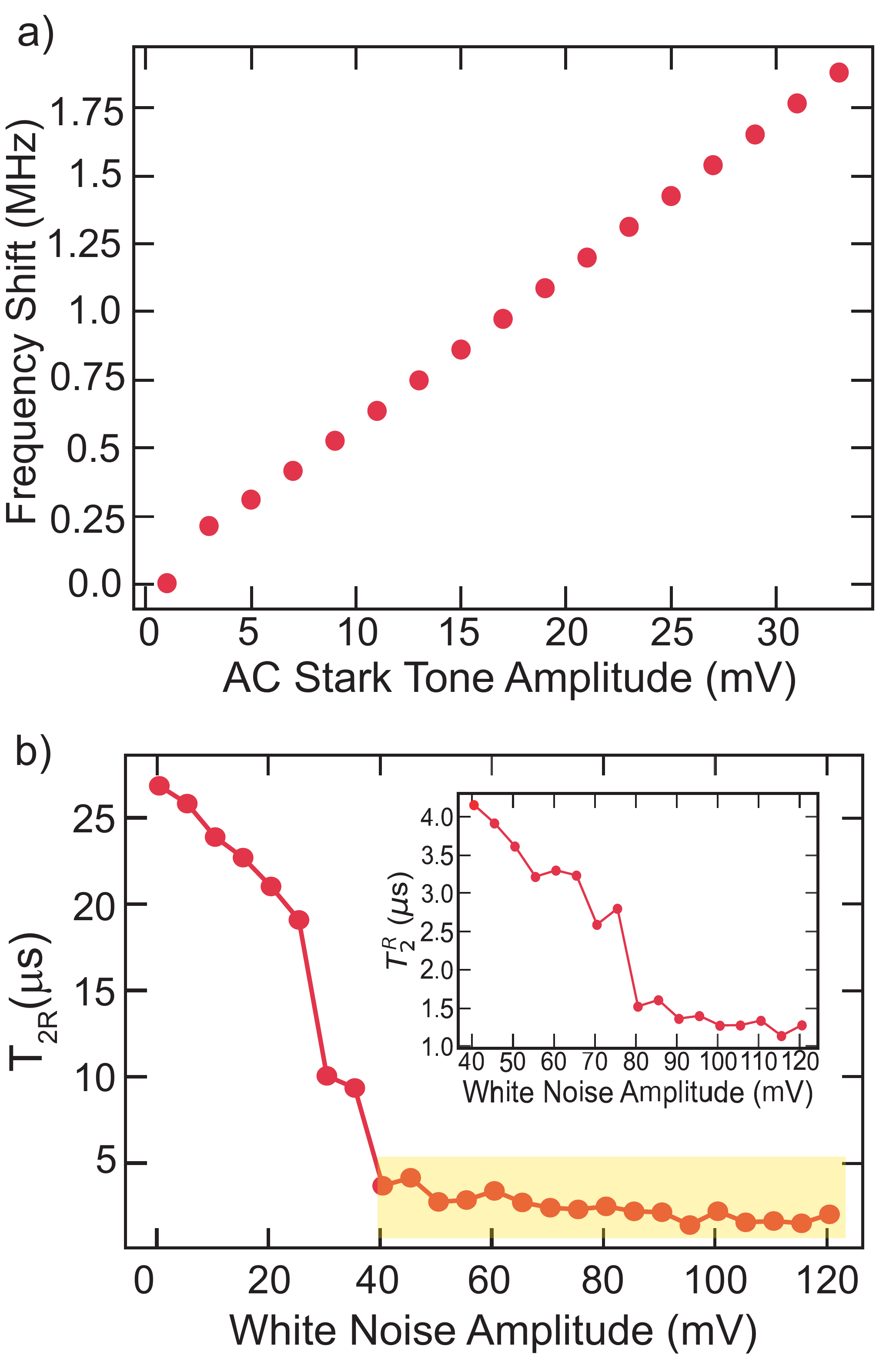}
    \caption{Calibration of white noise parameters. Plot (a) shows the linear relationship between the qubit frequency shift induced by the AC stark tone and AC stark tone amplitude $\mu$. Plot (b) shows how the coherence decreases with increasing white noise amplitude $\sigma$. Statistics for each amplitude point were generated by measuring coherence for 100 different white noise realizations. The inset corresponds to the yellow shaded region. The flatness at high amplitudes is due to $T_{max}\gg T_2^R$, where $T_{max}$ is the length of the Ramsey trace.}
    \label{fig:AC_stark_calib}
\end{figure}

The benefit of using the AC stark effect to inject white noise into the system is that there is a linear transfer between waveform amplitude and frequency change, and hence noise, induced at the qubit. The linear relation between qubit frequency shift and AC stark waveform amplitude is shown in Figure \ref{fig:AC_stark_calib}(a). The background coherence time $T_2^R=\tau_0$ is shown in Figure \ref{fig:AC_stark_calib}(b).

\subsection{Simulations}
\label{sec:Sims}
\subsubsection{Quantum Trajectory Simulation}
We choose the Stochastic Schr\"odinger Equation formalism for our simulations because it constitutes a straightforward and accurate method to simulate a single realization of a physical system coupled to its environment. 

Quantum trajectory techniques were developed by the field of Quantum Optics in the early 1990's to simulate dissipative dynamics \cite{OQS-Carmichael-book}. The main difference of such methods compared to the Master Equation formalism is that quantum trajectories can be used to describe \textit{single} realizations of an experiment on a quantum system instead of an ensemble of experimental realizations on a quantum system. Quantum trajectories have since proved a powerful tool in the physicists' arsenal, providing significant insights into the behavior of quantum systems, most importantly the century-old problem of wavefunction collapse, and tools for error correction such as measurement-based quantum feedback and control \cite{weber_2014,slichter_2011}. In classical dynamics, a trajectory describes the path an object takes in space. A quantum trajectory in contrast, describes how a quantum system evolves in the appropriate Hilbert space. A closed quantum system will evolve in a deterministic manner according to the Schr\"odinger equation. On the other hand, an open quantum system interacts with its environment by exchanging energy and information, and evolves stochastically. To generate a single quantum trajectory we solved 
\begin{align*}  
 \frac{d}{dt} \ket{\psi(t)} &=-i\hat{H} \ket{\psi(t)} \ ,\\
 \hat{H}&=\tfrac{1}{2}\omega_M(t)\sigma_i+\tfrac{1}{2}\omega_N(t)\sigma_j \ , \ 
 i\neq j \ ,
\end{align*}
where $\omega_{M,N}(t)$ are the Markovian and generalized Markovian noise realizations, respectively. The process was repeated for 1000 different noise realizations, and the results were averaged over these realizations to yield the ensemble dynamics. Because the Hamiltonian depends on timeseries that are evolving rapidly in time and have no closed-form solution, we used Matlab's non-stiff, variable-order differential equation solver 113. This solver determined the meshing of time automatically based on the solver tolerance criteria. We compared the efficiency and accuracy of this solver to those of other solvers such as solvers 45 and 78 and found that solver 113 performs the best in terms of speed and accuracy. 
\subsubsection{Simulation of higher transmon levels' effect on protocol}
\label{sec:anharmonicity}
\begin{figure}[h]
    \centering
    \includegraphics[width=0.85\columnwidth]{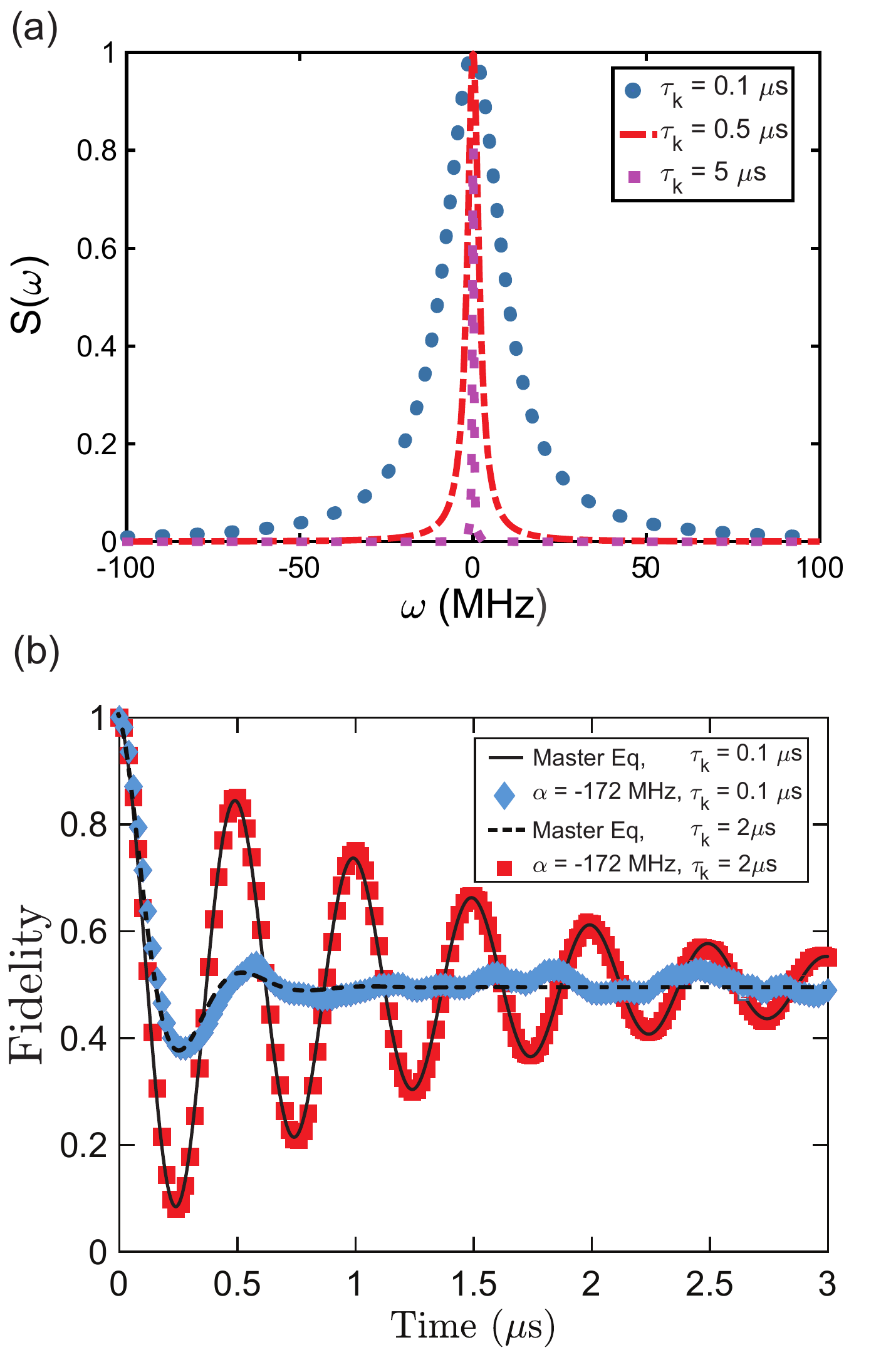}
    \caption{(a) Power spectrum of random telegraph noise for different switching rates. As the switching rate $1/\tau_k$ increases, the spectrum broadens and thus there is a higher chance of exciting the third energy level of the transmon. (b) Simulation of the effect of third energy level on protocol with $B_0 = 2$ MHz and $\alpha = -172$ MHz.}
    \label{fig:qutrit_sim_plot}
\end{figure}

In the prior simulations and the master equation solution, we treated the transmon as a true 2-state system (qubit). In reality our transmon is a weakly anharmonic oscillator with anharmonicity $\alpha \approx 2\pi \times -170$ MHz \cite{Koch_transmon}. For noise signals with high switching rates, i.e. small $\tau_k$, the noise bandwidth increases (see Fig. \ref{fig:qutrit_sim_plot}(a)), and so higher transitions may be accidentally driven. To test the validity of the 2-state approximation, we ran a set of simulations with the transmon treated as a 3-state system (qutrit). The Hamiltonian that governs the stochastic evolution of the wavefunction (in the qubit's rotating frame) is now given by:
\begin{equation}
    \hat{H}=\omega_M(t)\hat{a}^\dag a+\omega_N(t)(\hat{a}^{\dag}+\hat{a}) + \tfrac{\alpha}{2}\hat{a}^{\dag}\hat{a}(\hat{a}^{\dag}\hat{a}-\mathbb{I})
\end{equation}
where $\hat{a}^{\dag}$,$\hat{a}$ are the creation/annihilation operators truncated to the third level. 
We see no significant deviation from the exact master equation solution even for switching times as short as $\tau_k = 0.1$ $\mu$s. Our simulations thus show that our master equation emulation protocols are not signficantly affected by the finite anharmonicity of the transmon.
\subsubsection{Bloch-RedField Master Equation}
The Bloch-Redfield master equation (BRME) can be derived by starting with a generic system-bath interacting Hamiltonian, upon making a series of approximations, most notably, the Born and Markov approximations, but \textit{not} the secular (rotating wave) approximation, which would result in a master equation of the Lindblad form \cite{TOQS-book}. The BRME is convenient for describing a system coupled to a bath through a particular operator, where transition rates are determined by a power spectral density (PSD) function (though note, care must be taken in order to guarantee it defines a valid quantum process).

A white noise source corresponds to a flat PSD, which in the context of the present work where the environment is only coupled along a single axis, will induce dephasing at a constant rate (towards that axis), regardless of the qubit energy scale.

If however the PSD has structure, the dephasing rate will depend on the transition frequency ($\omega_1 - \omega_0$). It is common to introduce a high-frequency cutoff in the PSD, which will suppress dephasing between eigenstates that have a large energy gap.

In our work to understand the effect of a noise source which is not perfectly white, we introduce a flat power spectrum with an exponential tail, at some cutoff frequency $\omega_c$, i.e.
\begin{equation*}
    J(\omega) = \eta e^{-(\omega - \omega_c)\mathbf{1}_{\omega > \omega_c}} \ ,
\end{equation*}
where $\eta$ is a constant coupling strength, and the indicator function $\mathbf{1}_{\omega>\omega_c}$ is 1 for $\omega>\omega_c$ but otherwise 0. 

With this we can then simulate the equivalent set-up as described in the main text, with the white noise source replaced by one with a high frequency cutoff as above. We perform our simulations using QuTiP's \texttt{brmesolve} method \cite{qutip1, qutip2}. 
For example, in the XY protocol, the \texttt{a\_ops} parameter in QuTiP (which specifies the systems coupling with the environment) will be given by the $\sigma_y$ operator and a spectrum of the form $J(\omega)$ above. 

We pick $\eta=1/(2\tau_0)$, so that the dephasing rate without any system Hamiltonian ($\omega_0=\omega_1=0$) is identical in the white noise case and the case with $\omega_c < \infty$ (since here $J(0)=\eta$ is the only relevant quantity). The introduction of the telegraph Hamiltonian however changes the energy scale of the qubit, and therefore we can start to see differences in the decay, depending on the choice of $\omega_c$, an example of which is shown in Figure~\ref{fig:BW_plots}(c).

\bibliography{references.bib}

\end{document}